\PassOptionsToPackage{unicode}{hyperref}
\PassOptionsToPackage{hyphens}{url}
\documentclass[
  11pt,
]{article}
\usepackage{amsmath,amssymb}
\usepackage{iftex}
\ifPDFTeX
  \usepackage[T1]{fontenc}
  \usepackage[utf8]{inputenc}
  \usepackage{textcomp} 
\else 
  \usepackage{unicode-math} 
  \defaultfontfeatures{Scale=MatchLowercase}
  \defaultfontfeatures[\rmfamily]{Ligatures=TeX,Scale=1}
\fi
\usepackage{lmodern}
\ifPDFTeX\else
\fi
\IfFileExists{upquote.sty}{\usepackage{upquote}}{}
\IfFileExists{microtype.sty}{
  \usepackage[]{microtype}
  \UseMicrotypeSet[protrusion]{basicmath} 
}{}
\makeatletter
\@ifundefined{KOMAClassName}{
  \IfFileExists{parskip.sty}{%
    \usepackage{parskip}
  }{
    \setlength{\parindent}{0pt}
    \setlength{\parskip}{6pt plus 2pt minus 1pt}}
}{
  \KOMAoptions{parskip=half}}
\makeatother
\usepackage{xcolor}
\usepackage[margin=1in]{geometry}
\usepackage{color}
\usepackage{fancyvrb}

\DefineVerbatimEnvironment{Highlighting}{Verbatim}{commandchars=\\\{\}}
\usepackage{framed}
\definecolor{shadecolor}{RGB}{248,248,248}
\newenvironment{Shaded}{\begin{snugshade}}{\end{snugshade}}

\newcommand{\AttributeTok}[1]{\textcolor[rgb]{0.13,0.29,0.53}{#1}}

\newcommand{\CommentTok}[1]{\textcolor[rgb]{0.56,0.35,0.01}{\textit{#1}}}

\newcommand{\ConstantTok}[1]{\textcolor[rgb]{0.56,0.35,0.01}{#1}}

\newcommand{\DecValTok}[1]{\textcolor[rgb]{0.00,0.00,0.81}{#1}}

\newcommand{\FloatTok}[1]{\textcolor[rgb]{0.00,0.00,0.81}{#1}}
\newcommand{\FunctionTok}[1]{\textcolor[rgb]{0.13,0.29,0.53}{\textbf{#1}}}

\newcommand{\NormalTok}[1]{#1}

\newcommand{\OtherTok}[1]{\textcolor[rgb]{0.56,0.35,0.01}{#1}}

\newcommand{\SpecialCharTok}[1]{\textcolor[rgb]{0.81,0.36,0.00}{\textbf{#1}}}

\newcommand{\StringTok}[1]{\textcolor[rgb]{0.31,0.60,0.02}{#1}}

\usepackage{graphicx}
\makeatletter
\def\maxwidth{\ifdim\Gin@nat@width>\linewidth\linewidth\else\Gin@nat@width\fi}
\def\maxheight{\ifdim\Gin@nat@height>\textheight\textheight\else\Gin@nat@height\fi}
\makeatother
\setkeys{Gin}{width=\maxwidth,height=\maxheight,keepaspectratio}
\makeatletter
\def\fps@figure{htbp}
\makeatother
\providecommand{\tightlist}{%
  \setlength{\itemsep}{0pt}\setlength{\parskip}{0pt}}
\makeatletter
\def\verbatim{\normalsize\@verbatim \frenchspacing\@vobeyspaces \@xverbatim}
\makeatother

\usepackage{graphicx}

\usepackage{pdfpages}

\usepackage{dsfont}
\usepackage{longtable}
\usepackage{float}
\let\origfigure\figure
\let\endorigfigure\endfigure
\renewenvironment{figure}[1][2] {
    \expandafter\origfigure\expandafter[H]
} {
    \endorigfigure
}

\newcommand{\R}{\mathbb{R}}
\newcommand{\pkg}[1]{{\fontseries{b}\selectfont #1}}
\let\proglang=\textsf
\usepackage{mathrsfs}
\def\WT{\mathcal{W}}
\def\La{\mathcal L}

\def\D{D}
\def\W{W}
\def\R{\mathbb{R}}
\usepackage{booktabs}

\usepackage{booktabs}
\usepackage{longtable}
\usepackage{array}
\usepackage{multirow}
\usepackage{wrapfig}
\usepackage{float}
\usepackage{colortbl}
\usepackage{pdflscape}
\usepackage{tabu}
\usepackage{threeparttable}
\usepackage{threeparttablex}
\usepackage[normalem]{ulem}
\usepackage{makecell}
\usepackage{xcolor}
\ifLuaTeX
  \usepackage{selnolig}  
\fi
\usepackage[]{natbib}
\IfFileExists{bookmark.sty}{\usepackage{bookmark}}{\usepackage{hyperref}}
\IfFileExists{xurl.sty}{\usepackage{xurl}}{} 
\hypersetup{
  pdftitle={Gasper: GrAph Signal ProcEssing in R},
  pdfauthor={Basile de Loynes, Fabien Navarro, Baptiste Olivier},
  hidelinks,
  pdfcreator={LaTeX via pandoc}}

\title{Gasper: GrAph Signal ProcEssing in R}
\author{Basile de Loynes, Fabien Navarro, Baptiste Olivier}
\date{2023-12-28}

\begin{document}
\maketitle

\begin{abstract}
We present a short tutorial on to the use of the \proglang{R} \pkg{gasper} package. Gasper is a package dedicated to signal processing on graphs. It also provides an interface to the SuiteSparse Matrix Collection.
\end{abstract}

\hypertarget{introduction}{%
\section{Introduction}\label{introduction}}

The emerging field of Graph Signal Processing (GSP) aims to bridge the
gap between signal processing and spectral graph theory. One of the
objectives is to generalize fundamental analysis operations from regular
grid signals to irregular structures in the form of graphs. There is an
abundant literature on GSP, in particular we refer the reader to
\citet{shuman2013emerging} and \citet{ortega2018graph} for an
introduction to this field and an overview of recent developments,
challenges and applications. GSP has also given rise to numerous
applications in machine/deep learning: convolutional neural networks
(CNN) on graphs \citet{bruna2013spectral}, \citet{henaff2015deep},
\citet{defferrard2016convolutional}, semi-supervised classification with
graph CNN \citet{kipf2016semi}, \citet{hamilton2017inductive}, community
detection \citet{tremblay2014graph}, to name just a few.

Different software programs exist for processing signals on graphs, in
different languages. The Graph Signal Processing toolbox (GSPbox) is an
easy to use matlab toolbox that performs a wide variety of operations on
graphs. This toolbox was port to Python as the PyGSP
\citet{perraudin2014gspbox}. There is also another matlab toolbox the
Spectral Graph Wavelet Transform (SGWT) toolbox dedicated to the
implementation of the SGWT developed in \citet{hammond2011wavelets}.
However, to our knowledge, there are not yet any tools dedicated to GSP
in \proglang{R}. A development version of the \pkg{gasper} package is
currently available
online\footnote{\url{https://github.com/fabnavarro/gasper}}, while the
latest stable release can be obtained from the Comprehensive R Archive
Network\footnote{\url{https://cran.r-project.org/web/packages/gasper/}}.
In particular, it includes the methodology and
codes\footnote{\url{https://github.com/fabnavarro/SGWT-SURE}} developed
in \citet{de2019data} and provides an interface to the SuiteSparse
Matrix Collection \citet{davis2011university}.

This vignette is organized as follows. Section 2 introduces the
interface to the SuiteSparse Matrix Collection and some visualization
tools for GSP. Section 3 gives a short introduction to some underlying
concepts of GSP, focusing on the Graph Fourier Transform. Section 4
gives an illustration for denoising signals on graphs using SGWT,
thresholding techniques, and the minimization of Stein Unbiased Risk
Estimator for an automatic selection of the threshold parameter.

\hypertarget{graphs-collection-and-visualization}{%
\section{Graphs Collection and
Visualization}\label{graphs-collection-and-visualization}}

A certain number of graphs are present in the package. They are stored
as an Rdata file which contains a list consisting of the graph's weight
matrix \(W\) (in the form of a sparse matrix denoted by \texttt{sA}) and
the coordinates associated with the graph (if it has any).

An interface is also provided. It allows to retrieve the matrices
related to many problems provided by the SuiteSparse Matrix Collection
(formerly known as the University of Florida Sparse Matrix Collection)
\citet{davis2011university}, \citet{kolodziej19}. This collection is a
large and actively growing set of sparse matrices that arise in real
applications (as structural engineering, computational fluid dynamics,
computer graphics/vision, optimization, economic and financial modeling,
mathematics and statistics, to name just a few). For more details see
\url{https://sparse.tamu.edu/}.

The package includes the \texttt{SuiteSparseData} dataset, which
contains data from the SuiteSparse Matrix Collection. The structure of
this dataframe mirrors the structure of the main table presented on the
SuiteSparse Matrix Collection website, allowing users to query and
explore the dataset directly within \proglang{R}.

Here is a sample of the \texttt{SuiteSparseData} dataset, showing the
first 15 rows of the table:

\begin{Shaded}
\begin{Highlighting}[]
\FunctionTok{head}\NormalTok{(SuiteSparseData, }\DecValTok{15}\NormalTok{)}
\end{Highlighting}
\end{Shaded}

\begin{table}[H]

\caption{\label{tab:unnamed-chunk-2}Overview of the first 15 matrices from the SuiteSparse Matrix Collection.}
\centering
\begin{tabular}[t]{lllrrrll}
\toprule
ID & Name & Group & Rows & Cols & Nonzeros & Kind & Date\\
\midrule
\cellcolor{gray!6}{1} & \cellcolor{gray!6}{1138\_bus} & \cellcolor{gray!6}{HB} & \cellcolor{gray!6}{1138} & \cellcolor{gray!6}{1138} & \cellcolor{gray!6}{4054} & \cellcolor{gray!6}{Power Network Problem} & \cellcolor{gray!6}{1985}\\
2 & 494\_bus & HB & 494 & 494 & 1666 & Power Network Problem & 1985\\
\cellcolor{gray!6}{3} & \cellcolor{gray!6}{662\_bus} & \cellcolor{gray!6}{HB} & \cellcolor{gray!6}{662} & \cellcolor{gray!6}{662} & \cellcolor{gray!6}{2474} & \cellcolor{gray!6}{Power Network Problem} & \cellcolor{gray!6}{1985}\\
4 & 685\_bus & HB & 685 & 685 & 3249 & Power Network Problem & 1985\\
\cellcolor{gray!6}{5} & \cellcolor{gray!6}{abb313} & \cellcolor{gray!6}{HB} & \cellcolor{gray!6}{313} & \cellcolor{gray!6}{176} & \cellcolor{gray!6}{1557} & \cellcolor{gray!6}{Least Squares Problem} & \cellcolor{gray!6}{1974}\\
\addlinespace
6 & arc130 & HB & 130 & 130 & 1037 & Materials Problem & 1974\\
\cellcolor{gray!6}{7} & \cellcolor{gray!6}{ash219} & \cellcolor{gray!6}{HB} & \cellcolor{gray!6}{219} & \cellcolor{gray!6}{85} & \cellcolor{gray!6}{438} & \cellcolor{gray!6}{Least Squares Problem} & \cellcolor{gray!6}{1974}\\
8 & ash292 & HB & 292 & 292 & 2208 & Least Squares Problem & 1974\\
\cellcolor{gray!6}{9} & \cellcolor{gray!6}{ash331} & \cellcolor{gray!6}{HB} & \cellcolor{gray!6}{331} & \cellcolor{gray!6}{104} & \cellcolor{gray!6}{662} & \cellcolor{gray!6}{Least Squares Problem} & \cellcolor{gray!6}{1974}\\
10 & ash608 & HB & 608 & 188 & 1216 & Least Squares Problem & 1974\\
\addlinespace
\cellcolor{gray!6}{11} & \cellcolor{gray!6}{ash85} & \cellcolor{gray!6}{HB} & \cellcolor{gray!6}{85} & \cellcolor{gray!6}{85} & \cellcolor{gray!6}{523} & \cellcolor{gray!6}{Least Squares Problem} & \cellcolor{gray!6}{1974}\\
12 & ash958 & HB & 958 & 292 & 1916 & Least Squares Problem & 1974\\
\cellcolor{gray!6}{13} & \cellcolor{gray!6}{bcspwr01} & \cellcolor{gray!6}{HB} & \cellcolor{gray!6}{39} & \cellcolor{gray!6}{39} & \cellcolor{gray!6}{131} & \cellcolor{gray!6}{Power Network Problem} & \cellcolor{gray!6}{1981}\\
14 & bcspwr02 & HB & 49 & 49 & 167 & Power Network Problem & 1981\\
\cellcolor{gray!6}{15} & \cellcolor{gray!6}{bcspwr03} & \cellcolor{gray!6}{HB} & \cellcolor{gray!6}{118} & \cellcolor{gray!6}{118} & \cellcolor{gray!6}{476} & \cellcolor{gray!6}{Power Network Problem} & \cellcolor{gray!6}{1981}\\
\bottomrule
\end{tabular}
\end{table}

For example, to retrieve all undirected graphs with between 100 and 150
columns and rows:

\begin{Shaded}
\begin{Highlighting}[]
\NormalTok{filtered\_mat }\OtherTok{\textless{}{-}}\NormalTok{ SuiteSparseData[SuiteSparseData}\SpecialCharTok{$}\NormalTok{Kind }\SpecialCharTok{==} \StringTok{"Undirected Graph"} \SpecialCharTok{\&} 
\NormalTok{                 SuiteSparseData}\SpecialCharTok{$}\NormalTok{Rows }\SpecialCharTok{\textgreater{}=} \DecValTok{100} \SpecialCharTok{\&}\NormalTok{ SuiteSparseData}\SpecialCharTok{$}\NormalTok{Rows }\SpecialCharTok{\textless{}=} \DecValTok{150} \SpecialCharTok{\&}
\NormalTok{                 SuiteSparseData}\SpecialCharTok{$}\NormalTok{Cols }\SpecialCharTok{\textgreater{}=} \DecValTok{100} \SpecialCharTok{\&}\NormalTok{ SuiteSparseData}\SpecialCharTok{$}\NormalTok{Cols }\SpecialCharTok{\textless{}=} \DecValTok{150}\NormalTok{, ]}
\NormalTok{filtered\_mat}
\end{Highlighting}
\end{Shaded}

\begin{table}[H]

\caption{\label{tab:unnamed-chunk-4}Subset of undirected matrices with 100 to 150 rows and columns.}
\centering
\begin{tabular}[t]{llllrrrll}
\toprule
  & ID & Name & Group & Rows & Cols & Nonzeros & Kind & Date\\
\midrule
\cellcolor{gray!6}{1484} & \cellcolor{gray!6}{1484} & \cellcolor{gray!6}{GD06\_theory} & \cellcolor{gray!6}{Pajek} & \cellcolor{gray!6}{101} & \cellcolor{gray!6}{101} & \cellcolor{gray!6}{380} & \cellcolor{gray!6}{Undirected Graph} & \cellcolor{gray!6}{2006}\\
1497 & 1497 & GD98\_c & Pajek & 112 & 112 & 336 & Undirected Graph & 1998\\
\cellcolor{gray!6}{2389} & \cellcolor{gray!6}{2389} & \cellcolor{gray!6}{adjnoun} & \cellcolor{gray!6}{Newman} & \cellcolor{gray!6}{112} & \cellcolor{gray!6}{112} & \cellcolor{gray!6}{850} & \cellcolor{gray!6}{Undirected Graph} & \cellcolor{gray!6}{2006}\\
2403 & 2403 & polbooks & Newman & 105 & 105 & 882 & Undirected Graph & 2001\\
\bottomrule
\end{tabular}
\end{table}

The \texttt{download\_graph} function allows to download a matrix from
this collection, based on the name of the matrix and the name of the
group that provides it. An example is given below

\begin{Shaded}
\begin{Highlighting}[]
\NormalTok{matrixname }\OtherTok{\textless{}{-}} \StringTok{"grid1"}
\NormalTok{groupname }\OtherTok{\textless{}{-}} \StringTok{"AG{-}Monien"}
\FunctionTok{download\_graph}\NormalTok{(matrixname, groupname)}
\end{Highlighting}
\end{Shaded}

\begin{Shaded}
\begin{Highlighting}[]
\FunctionTok{attributes}\NormalTok{(grid1)}
\CommentTok{\#\textgreater{} $names}
\CommentTok{\#\textgreater{} [1] "sA"   "xy"   "dim"  "temp"}
\end{Highlighting}
\end{Shaded}

The output is stored (in a temporary folder) as a list composed of:

\begin{itemize}
\tightlist
\item
  \texttt{sA} the corresponding sparse matrix (in compressed sparse
  column format);
\end{itemize}

\begin{Shaded}
\begin{Highlighting}[]
\FunctionTok{str}\NormalTok{(grid1}\SpecialCharTok{$}\NormalTok{sA)}
\CommentTok{\#\textgreater{} Formal class \textquotesingle{}dsCMatrix\textquotesingle{} [package "Matrix"] with 7 slots}
\CommentTok{\#\textgreater{}   ..@ i       : int [1:476] 173 174 176 70 71 74 74 75 77 77 ...}
\CommentTok{\#\textgreater{}   ..@ p       : int [1:253] 0 3 6 9 12 15 18 21 24 27 ...}
\CommentTok{\#\textgreater{}   ..@ Dim     : int [1:2] 252 252}
\CommentTok{\#\textgreater{}   ..@ Dimnames:List of 2}
\CommentTok{\#\textgreater{}   .. ..$ : NULL}
\CommentTok{\#\textgreater{}   .. ..$ : NULL}
\CommentTok{\#\textgreater{}   ..@ x       : num [1:476] 1 1 1 1 1 1 1 1 1 1 ...}
\CommentTok{\#\textgreater{}   ..@ uplo    : chr "L"}
\CommentTok{\#\textgreater{}   ..@ factors : list()}
\end{Highlighting}
\end{Shaded}

\begin{itemize}
\tightlist
\item
  possibly coordinates \texttt{xy} (stored in a \texttt{data.frame});
\end{itemize}

\begin{Shaded}
\begin{Highlighting}[]
\FunctionTok{head}\NormalTok{(grid1}\SpecialCharTok{$}\NormalTok{xy, }\DecValTok{3}\NormalTok{)}
\CommentTok{\#\textgreater{}            x       y}
\CommentTok{\#\textgreater{} [1,] 0.00000 0.00000}
\CommentTok{\#\textgreater{} [2,] 2.88763 3.85355}
\CommentTok{\#\textgreater{} [3,] 3.14645 4.11237}
\end{Highlighting}
\end{Shaded}

\begin{itemize}
\tightlist
\item
  \texttt{dim"} the numbers of rows, columns and numerically nonzero
  elements;
\end{itemize}

\begin{Shaded}
\begin{Highlighting}[]
\NormalTok{grid1}\SpecialCharTok{$}\NormalTok{dim}
\CommentTok{\#\textgreater{}   NumRows NumCols NonZeros}
\CommentTok{\#\textgreater{} 1     252     252      476}
\end{Highlighting}
\end{Shaded}

\begin{itemize}
\tightlist
\item
  \texttt{temp} the path to the temporary directory where the matrix and
  downloaded files (including singular values if requested) are stored.
\end{itemize}

\begin{Shaded}
\begin{Highlighting}[]
\FunctionTok{list.files}\NormalTok{(grid1}\SpecialCharTok{$}\NormalTok{temp)}
\end{Highlighting}
\end{Shaded}

Metadata associated with the matrix can be display via

\begin{Shaded}
\begin{Highlighting}[]
\FunctionTok{file.show}\NormalTok{(}\FunctionTok{paste}\NormalTok{(grid1}\SpecialCharTok{$}\NormalTok{temp,}\StringTok{"grid1"}\NormalTok{,}\AttributeTok{sep=}\StringTok{""}\NormalTok{))}
\end{Highlighting}
\end{Shaded}

or in the console:

\begin{Shaded}
\begin{Highlighting}[]
\FunctionTok{cat}\NormalTok{(}\FunctionTok{readLines}\NormalTok{(}\FunctionTok{paste}\NormalTok{(grid1}\SpecialCharTok{$}\NormalTok{temp,}\StringTok{"grid1"}\NormalTok{,}\AttributeTok{sep=}\StringTok{""}\NormalTok{), }\AttributeTok{n=}\DecValTok{14}\NormalTok{), }\AttributeTok{sep =} \StringTok{"}\SpecialCharTok{\textbackslash{}n}\StringTok{"}\NormalTok{)}
\end{Highlighting}
\end{Shaded}

\texttt{download\_graph} function has an optional \texttt{svd} argument;
setting \texttt{svd\ =\ "TRUE"} downloads a ``.mat'' file containing the
singular values of the matrix, if available.

For further insights, the \texttt{get\_graph\_info} function retrieve
detailed information about the matrix from the SuiteSparse Matrix
Collection website. \texttt{get\_graph\_info} fetches the three tables
with ``MatrixInformation'', ``MatrixProperties,'' and ``SVDStatistics'',
providing a comprehensive overview of the matrix (\texttt{rvest} package
needs to be installed).

\begin{Shaded}
\begin{Highlighting}[]
\NormalTok{matrix\_info }\OtherTok{\textless{}{-}} \FunctionTok{get\_graph\_info}\NormalTok{(matrixname, groupname)}
\NormalTok{matrix\_info}
\end{Highlighting}
\end{Shaded}

The \texttt{download\_graph} function also has an optional argument
\texttt{add\_info} which, when set to \texttt{TRUE}, automatically calls
\texttt{get\_graph\_info} and appends the retrieved information to the
output of \texttt{download\_graph}. This makes it easy to get both the
graph data and its associated information in a single function call.

\begin{Shaded}
\begin{Highlighting}[]
\NormalTok{downloaded\_graph }\OtherTok{\textless{}{-}} \FunctionTok{download\_graph}\NormalTok{(matrixname, groupname, }\AttributeTok{add\_info =} \ConstantTok{TRUE}\NormalTok{)}
\NormalTok{downloaded\_graph}\SpecialCharTok{$}\NormalTok{info}
\end{Highlighting}
\end{Shaded}

\begin{table}[H]
\caption{\label{tab:unnamed-chunk-16}Matrix Information (left) and Matrix Properties (right).}

\begin{tabular}[t]{ll}
\toprule
  & MatrixInformation\\
\midrule
Name & grid1\\
Group & AG-Monien\\
Matrix ID & 2416\\
Num Rows & 252\\
Num Cols & 252\\
\addlinespace
Nonzeros & 952\\
Pattern Entries & 952\\
Kind & 2D/3D Problem\\
Symmetric & Yes\\
Date & 1998\\
\addlinespace
Author & R. Diekmann, R. Preis\\
Editor & R. Diekmann, R. Preis\\
\bottomrule
\end{tabular}
\begin{tabular}[t]{ll}
\toprule
  & SVDStatistics\\
\midrule
Structural Rank & 252\\
Structural Rank Full & true\\
Num Dmperm Blocks & 2\\
Strongly Connect Components & 1\\
Num Explicit Zeros & 0\\
\addlinespace
Pattern Symmetry & 100\%\\
Numeric Symmetry & 100\%\\
Cholesky Candidate & no\\
Positive Definite & no\\
Type & binary\\
\bottomrule
\end{tabular}
\end{table}

The package also allows to plot a (planar) graph using the function
\texttt{plot\_graph}. It also contains a function to plot signals
defined on top of the graph \texttt{plot\_signal}.

\begin{Shaded}
\begin{Highlighting}[]
\NormalTok{f }\OtherTok{\textless{}{-}} \FunctionTok{rnorm}\NormalTok{(}\FunctionTok{nrow}\NormalTok{(grid1}\SpecialCharTok{$}\NormalTok{sA))}
\FunctionTok{plot\_graph}\NormalTok{(grid1)}
\FunctionTok{plot\_signal}\NormalTok{(grid1, f, }\AttributeTok{size =} \DecValTok{2}\NormalTok{)}
\end{Highlighting}
\end{Shaded}

\begin{figure}

{\centering \includegraphics{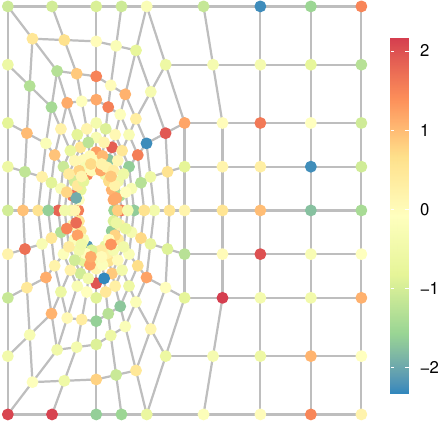} 

}

\caption{Graph (left) and graph signal (right).}\label{fig:unnamed-chunk-17}
\end{figure}

In cases where these coordinates are not available, \texttt{plot\_graph}
employs simple spectral graph drawing to calculate some node
coordinates. This is done using the function \texttt{spectral\_coords},
which computes the spectral coordinates based on the eigenvectors
associated with the two smallest non-zero eigenvalues of the graph's
Laplacian \citet{hall70}.

\hypertarget{a-short-introduction-to-graph-signal-processing}{%
\section{A Short Introduction to Graph Signal
Processing}\label{a-short-introduction-to-graph-signal-processing}}

Graph theory provides a robust mathematical framework for representing
complex systems. In this context, entities are modeled as vertices (or
nodes) and their interconnections as edges, encapsulating a broad
spectrum of real-world phenomena from social and communication networks
to molecular structures and brain connectivity patterns.

Among the diverse types of graphs, such as undirected, directed,
weighted, bipartite, and multigraphs, each offers distinct analytical
advantages tailored to specific contexts. This vignette is devoted to
undirected, connected, graphs where edges link two vertices
symmetrically, often with weighted values to express connection strength
or intensity. For instance, in a road network graph, the weights might
correspond to the length of each road segment.

Graphs are defined by \({G}=({V}, {E})\), where \({V}\) denotes the set
of vertices or nodes and \({E}\) represents the set of edges. Each edge
\((i, j) \in {E}\) connects nodes \(i\) and \(j\), potentially with an
associated weight \(w_{ij}\). The connectivity and interaction structure
of \({G}\) is encoded in the adjacency matrix \(W\), where
\(w_{ij} = w_{ji}\) for \(i,j\in V\). The size of the graph is the
number of nodes \(n=|V|\). The degree matrix \(\D\) is a diagonal matrix
with \(\D_{ii} = \sum_{j\in V} w_{ij}\). These matrices, \(\W\) and
\(\D\), serve as the foundation for analyzing signal behavior on graph
structures in GSP.

The spectral properties of the Laplacian matrices offer deep insights
into the structure of graphs. The unnormalized Laplacian,
\(\La = \D - \W\), has non-negative eigenvalues, with the smallest being
zero, indicating the number of connected components in the graph. This
number can be retrieved from the ``MatrixProperty'' dataframe using the
\texttt{get\_graph\_info} function, if the graph has been downloaded
with \texttt{download\_graph}, or computed using Depth-First Search
algorithm, using \pkg{igraph} \proglang{R} package for instance
\cite{igraph}.

\begin{Shaded}
\begin{Highlighting}[]
\NormalTok{grid1}\SpecialCharTok{$}\NormalTok{info}\SpecialCharTok{$}\NormalTok{MatrixProp[}\StringTok{"Strongly Connect Components"}\NormalTok{,]}
\CommentTok{\#\textgreater{} [1] "1"}
\end{Highlighting}
\end{Shaded}

On the other hand, the normalized Laplacian,
\(\La_{\mathrm{norm}} = I - \D^{-1/2} \W \D^{-1/2}\), has a spectrum
that typically lies between 0 and 2. The zero eigenvalue corresponds to
the number of connected components, and the first non-zero smallest
eigenvalue, often referred to as the spectral gap, plays a crucial role
in determining the graph's propensity for clustering. The larger this
gap, the more pronounced the cluster structures within the graph. By
scaling the eigenvalues according to node degrees, the normalized
Laplacian accentuates the separation between clusters, making the
spectral gap a significant measure in spectral graph clustering
algorithms. However, a large spectral gap might present challenges for
fast spectral filtering, especially depending on the approximation
methods used, where it could lead to issues in convergence or
computational efficiency.

Lastly, the random walk Laplacian,
\(\La_{\mathrm{rw}} = I - \D^{-1}\W\), is generally better suited for
directed graphs and scenarios involving random walk dynamics. Its
spectrum is also non-negative. The choice of which Laplacian to use is
dictated by the particular graph properties one aims to emphasize or
analyze in a given application.

The \texttt{laplacian\_mat} function (which supports both standard and
sparse matrix representations) allows to compute those three forms of
Laplacian matrices. For example, let \(G=({V}, {E})\) a simple
undirected graph with the vertex set \({V} = \{1, 2, 3\}\) and the edge
set \({E} = \{\{1, 2\}, \{2, 3\}\}\). The corresponding adjacency matrix
\(\W\), as well as its unnormalized, normalized, and random walk
Laplacians, can be represented and calculated as follows:

\begin{Shaded}
\begin{Highlighting}[]
\NormalTok{W }\OtherTok{\textless{}{-}} \FunctionTok{matrix}\NormalTok{(}\FunctionTok{c}\NormalTok{(}\DecValTok{0}\NormalTok{, }\DecValTok{1}\NormalTok{, }\DecValTok{0}\NormalTok{,}
              \DecValTok{1}\NormalTok{, }\DecValTok{0}\NormalTok{, }\DecValTok{1}\NormalTok{,}
              \DecValTok{0}\NormalTok{, }\DecValTok{1}\NormalTok{, }\DecValTok{0}\NormalTok{), }\AttributeTok{ncol=}\DecValTok{3}\NormalTok{)}
\FunctionTok{laplacian\_mat}\NormalTok{(W, }\StringTok{"unnormalized"}\NormalTok{)}
\CommentTok{\#\textgreater{}      [,1] [,2] [,3]}
\CommentTok{\#\textgreater{} [1,]    1   {-}1    0}
\CommentTok{\#\textgreater{} [2,]   {-}1    2   {-}1}
\CommentTok{\#\textgreater{} [3,]    0   {-}1    1}
\FunctionTok{laplacian\_mat}\NormalTok{(W, }\StringTok{"normalized"}\NormalTok{)}
\CommentTok{\#\textgreater{}            [,1]       [,2]       [,3]}
\CommentTok{\#\textgreater{} [1,]  1.0000000 {-}0.7071068  0.0000000}
\CommentTok{\#\textgreater{} [2,] {-}0.7071068  1.0000000 {-}0.7071068}
\CommentTok{\#\textgreater{} [3,]  0.0000000 {-}0.7071068  1.0000000}
\FunctionTok{laplacian\_mat}\NormalTok{(W, }\StringTok{"randomwalk"}\NormalTok{)}
\CommentTok{\#\textgreater{}      [,1] [,2] [,3]}
\CommentTok{\#\textgreater{} [1,]  1.0   {-}1  0.0}
\CommentTok{\#\textgreater{} [2,] {-}0.5    1 {-}0.5}
\CommentTok{\#\textgreater{} [3,]  0.0   {-}1  1.0}
\end{Highlighting}
\end{Shaded}

GSP extends classical signal processing concepts to signals defined on
graphs. Let \({G}\) be a graph on the vertex set
\({V} = \{v_1, \ldots, v_n\}\). A graph signal on \({G}\) is a function
\(f : {V} \rightarrow \R\), that assigns a value to each node of the
graph. It can be represented as a vector
\((f(v_1), f(v_2), \ldots, f(v_n))^\top \in \R^n\), where each entry
\(f_i\) corresponds to the signal value at node \(i\).

The Laplacian quadratic form \(f^\top\La f\) gives a measure of a graph
signal's smoothness: \[ 
f^\top \La f = \frac{1}{2} \sum_{(i,j) \in E} w_{ij} (f_i - f_j)^2,
\] where a lower value suggests that the signal varies little between
connected nodes, and thus is smoother on the graph. It's a global
measure of the graph's ``frequency'' content which is insightful for
understanding the overall variation in graph signals. The
\texttt{smoothmodulus} function calculates this form for a given graph
signal, returning a scalar value that quantifies the signal's smoothness
in relation to the graph's structure. Moreover, the \texttt{randsignal}
function can be used to generate graph signals with varying smoothness
properties.

To analyze graph signals, the concept of the Graph Fourier Transform
(GFT) is fundamental. The GFT provides a means to represent graph
signals in the frequency domain, analogous to the classical Fourier
Transform for traditional signals. Given a graph \({G}\), a GFT can be
defined as the representation of signals on an orthonormal basis for
\(\R^n\) consisting of eigenvectors of the graph shift operator. The
choice of graph shift operator is essential, as it determines the basis
for the GFT, it can be either the Laplacian matrix or the adjacency
matrix. In this tutorial, we primarily focus on signal processing using
the Laplacian matrix as the shift operator.

For undirected graphs, the Laplacian matrix \(\La\) is symmetric and
positive semi-definite, with non-negative real eigenvalues. Given the
eigenvalue decomposition of the graph Laplacian \(\La = U \Lambda U^T\),
where \(U\) is the matrix of eigenvectors and \(\Lambda\) is the
diagonal matrix of eigenvalues, the GFT of a signal \(f\) is given by
\(\hat{f} = U^T f\). Here, \(\hat{f}\) represents the graph signal in
the frequency domain. The elements of \(\hat{f}\) are the coefficients
of the signal \(f\) with respect to the eigenvectors of \(\La\), which
can be interpreted as the frequency components of the signal on the
graph.

The inverse GFT is given by \(f = U \hat{f}\). This allows for the
reconstruction of the graph signal in the vertex domain from its
frequency representation. The GFT provides a powerful tool for analyzing
and processing signals on graphs. It enables the identification of
signal components that vary smoothly or abruptly over the graph,
facilitating tasks such as filtering, denoising, and compression of
graph signals. The function \texttt{forward\_gft} allows to perform a
GFT decomposition and to obtain the associated Fourier coefficients. The
function \texttt{inverse\_gft} allows to make the reconstruction.

\hypertarget{data-driven-graph-signal-denoising}{%
\section{Data-Driven Graph Signal
Denoising}\label{data-driven-graph-signal-denoising}}

We give an example of an application in the case of the denoising of a
noisy signal \(f\) defined on a graph \(G\) with set of vertices \(V\).
More precisely, the (unnormalized) graph Laplacian matrix
\(\La\in\R^{V\times V}\) associated with \(G\) is the symmetric matrix
defined as \(\La=\D - \W\), where \(\W\) is the matrix of weights with
coefficients \((w_{ij})_{i,j\in V}\), and \(\D\) the diagonal matrix
with diagonal coefficients \(\D_{ii}= \sum_{j\in V} w_{ij}\). A signal
\(f\) on the graph \(G\) is a function \(f:V\rightarrow \R\).

The degradation model can be written as \[
\tilde f = f + \xi,
\] where \(\xi\sim\mathcal{N}(0,\sigma^2)\). The purpose of denoising is
to build an estimator of \(f\) that depends only on \(\tilde f\).

A simple way to construct an effective non-linear estimator is obtained
by thresholding the SGWT coefficients of \(f\) on a frame (see
\citet{hammond2011wavelets} for details about the SGWT).

A general thresholding operator \(\tau\) with threshold parameter
\(t\geq 0\) applied to some signal \(f\) is defined as
\begin{equation}\label{eq:tau}
\tau(x,t)=x\max \{ 1-t^{\beta}|x|^{-\beta},0 \},
\end{equation} with \(\beta \geq 1\). The most popular choices are the
soft thresholding (\(\beta=1\)), the James-Stein thresholding
(\(\beta=2\)) and the hard thresholding (\(\beta=\infty\)).

Given the Laplacian and a given frame, denoising in this framework can
be summarized as follows:

\begin{itemize}
\item
  Analysis: compute the SGWT transform \(\WT \tilde f\);
\item
  Thresholding: apply a given thresholding operator to the coefficients
  \(\WT \tilde f\);
\item
  Synthesis: apply the inverse SGWT transform to obtain an estimation of
  the original signal.
\end{itemize}

Each of these steps can be performed via one of the functions
\texttt{analysis}, \texttt{synthesis}, \texttt{beta\_thresh}. Laplacian
is given by the function \texttt{laplacian\_mat}. The
\texttt{tight\_frame} function allows the construction of a tight frame
based on \citet{gobel2018construction} and \citet{coulhon2012heat}. In
order to select a threshold value, we consider the method developed in
\citet{de2019data} which consists in determining the threshold that
minimizes the Stein unbiased risk estimator (SURE) in a graph setting
(see \citet{de2019data} for more details).

We give an illustrative example on the \texttt{grid1} graph from the
previous section. We start by calculating, the Laplacian matrix (from
the adjacency matrix), its eigendecomposition and the frame
coefficients.

\begin{Shaded}
\begin{Highlighting}[]
\NormalTok{A }\OtherTok{\textless{}{-}}\NormalTok{ grid1}\SpecialCharTok{$}\NormalTok{sA}
\NormalTok{L }\OtherTok{\textless{}{-}} \FunctionTok{laplacian\_mat}\NormalTok{(A)}
\NormalTok{val1 }\OtherTok{\textless{}{-}} \FunctionTok{eigensort}\NormalTok{(L)}
\NormalTok{evalues }\OtherTok{\textless{}{-}}\NormalTok{ val1}\SpecialCharTok{$}\NormalTok{evalues}
\NormalTok{evectors }\OtherTok{\textless{}{-}}\NormalTok{ val1}\SpecialCharTok{$}\NormalTok{evectors}
\CommentTok{\#{-} largest eigenvalue}
\NormalTok{lmax }\OtherTok{\textless{}{-}} \FunctionTok{max}\NormalTok{(evalues)}
\CommentTok{\#{-} parameter that controls the scale number}
\NormalTok{b }\OtherTok{\textless{}{-}} \DecValTok{2}
\NormalTok{tf }\OtherTok{\textless{}{-}} \FunctionTok{tight\_frame}\NormalTok{(evalues, evectors, }\AttributeTok{b=}\NormalTok{b)}
\end{Highlighting}
\end{Shaded}

Wavelet frames can be seen as special filter banks. The tight-frame
considered here is a finite collection \((\psi_j)_{j=0, \ldots,J}\)
forming a finite partition of unity on the compact \([0,\lambda_1]\),
where \(\lambda_1\) is the largest eigenvalue of the Laplacian spectrum
\(\mathrm{sp}(\La)\). This partition is defined as follows: let
\(\omega : \mathbb R^+ \rightarrow [0,1]\) be some function with support
in \([0,1]\), satisfying \(\omega \equiv 1\) on \([0,b^{-1}]\), for some
\(b>1\), and set \begin{equation*}
\psi_0(x)=\omega(x)~~\textrm{and}~~\psi_j(x)=\omega(b^{-j}x)-\omega(b^{-j+1}x)~~\textrm{for}~~j=1, \ldots, J,~~\textrm{where}~~J= \left \lfloor \frac{\log \lambda_1}{\log b} \right \rfloor + 2.
\end{equation*} Thanks to Parseval's identity, the following set of
vectors is a tight frame: \[
\mathfrak F = \left \{ \sqrt{\psi_j}(\La)\delta_i, j=0, \ldots, J, i \in V \right \}.
\] The \texttt{plot\_filter} function allows to represent the elements
\(\sqrt{\psi_j}\) (filters) of this partition, with \[
\omega(x) = \begin{cases}
1 & \text{if } x \in [0,b^{-1}] \\
b \cdot \frac{x}{1 - b} + \frac{b}{b - 1} & \text{if } x \in (b^{-1}, 1] \\
0 & \text{if } x > 1
\end{cases}
\] which corresponds to the tigth-frame constructed from the
\texttt{zetav} function.

\begin{Shaded}
\begin{Highlighting}[]
\FunctionTok{plot\_filter}\NormalTok{(lmax,b)}
\end{Highlighting}
\end{Shaded}

\begin{figure}

{\centering \includegraphics{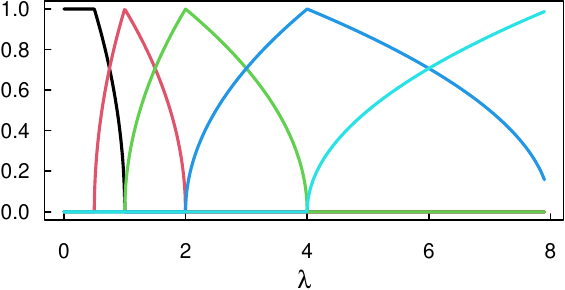} 

}

\caption{Plot of the spectral graph filters on the spectrum of grid1 graph.}\label{fig:unnamed-chunk-23}
\end{figure}

The SGWT of a signal \(f \in \mathbb R^V\) is given by \[
\WT f = \left ( \sqrt{\psi_0}(\La)f^{T},\ldots,\sqrt{\psi_J}(\La)f^{T} \right )^{T} \in \mathbb R^{n(J+1)}.
\] The adjoint linear transformation \(\WT^\ast\) of \(\WT\) is: \[
\WT^\ast \left (\eta_{0}^{T}, \eta_{1}^{T}, \ldots, \eta_{J}^T \right )^{T} = \sum_{j\geq 0} \sqrt{\psi_j}(\La)\eta_{j}.
\] The tightness of the underlying frame implies that
\(\WT^\ast \WT=\mathrm{Id}_{\mathbb R^V}\) so that a signal
\(f \in \mathbb R^V\) can be recovered by applying \(\WT^\ast\) to its
wavelet coefficients
\(((\WT f)_i)_{i=1, \ldots, n(J+1)} \in \mathbb R^{n(J+1)}\).

Then, noisy observations \(\tilde f\) are generated from a random signal
\(f\).

\begin{Shaded}
\begin{Highlighting}[]
\NormalTok{n }\OtherTok{\textless{}{-}} \FunctionTok{nrow}\NormalTok{(L)}
\NormalTok{f }\OtherTok{\textless{}{-}} \FunctionTok{randsignal}\NormalTok{(}\FloatTok{0.01}\NormalTok{, }\DecValTok{3}\NormalTok{, A)}
\NormalTok{sigma }\OtherTok{\textless{}{-}} \FloatTok{0.01}
\NormalTok{noise }\OtherTok{\textless{}{-}} \FunctionTok{rnorm}\NormalTok{(n, }\AttributeTok{sd =}\NormalTok{ sigma)}
\NormalTok{tilde\_f }\OtherTok{\textless{}{-}}\NormalTok{ f }\SpecialCharTok{+}\NormalTok{ noise}
\end{Highlighting}
\end{Shaded}

Below is a graphical representation of the original signal and its noisy
version.

\begin{Shaded}
\begin{Highlighting}[]
\FunctionTok{plot\_signal}\NormalTok{(grid1, f, }\AttributeTok{size =} \DecValTok{2}\NormalTok{)}
\FunctionTok{plot\_signal}\NormalTok{(grid1, tilde\_f, }\AttributeTok{size =} \DecValTok{2}\NormalTok{)}
\end{Highlighting}
\end{Shaded}

\begin{figure}

{\centering \includegraphics{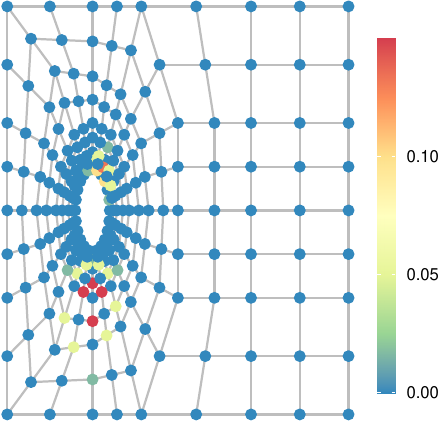} \includegraphics{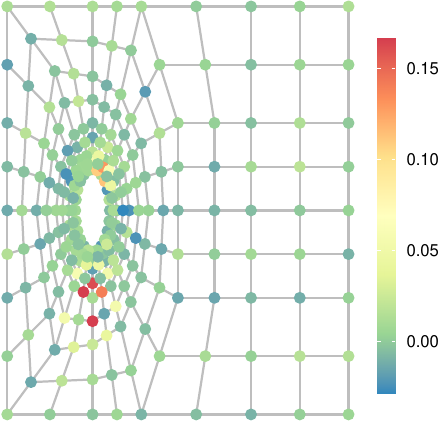} 

}

\caption{Original graph signal (left) and noisy observations (right).}\label{fig:unnamed-chunk-25}
\end{figure}

We compute the SGWT transforms \(\WT \tilde f\) and \(\WT f\).

\begin{Shaded}
\begin{Highlighting}[]
\NormalTok{wcn }\OtherTok{\textless{}{-}} \FunctionTok{analysis}\NormalTok{(tilde\_f,tf)}
\NormalTok{wcf }\OtherTok{\textless{}{-}} \FunctionTok{analysis}\NormalTok{(f,tf)}
\end{Highlighting}
\end{Shaded}

An alternative to avoid frame calculation is given by the
\texttt{forward\_sgwt} function which provides a forward SGWT. For
example:

\begin{Shaded}
\begin{Highlighting}[]
\NormalTok{wcf }\OtherTok{\textless{}{-}} \FunctionTok{forward\_sgwt}\NormalTok{(f, evalues, evectors, }\AttributeTok{b=}\NormalTok{b)}
\end{Highlighting}
\end{Shaded}

The optimal threshold is then determined by minimizing the SURE (using
Donoho and Johnstone's trick \citet{donoho1995adapting} which remains
valid here, see \citet{de2019data}). More precisely, the SURE for a
general thresholding process \(h\) is given by the following identity\\
\begin{equation}
\mathbf{SURE}(h)=-n \sigma^2 + \|h(\widetilde F)-\widetilde F\|^2 + 2 \sum_{i,j=1}^{n(J+1)} \gamma_{i,j}^2 \partial_j h_i(\widetilde F),
\end{equation} where \(\gamma_{i,j}^2=\sigma^2(\WT \WT ^\ast)_{i,j}\)
that can be computed from the frame (or estimated via Monte-Carlo
simulation). The \texttt{SURE\_thresh}/\texttt{SURE\_MSEthresh} allow to
evaluate the SURE (in a global fashion) considering the general
thresholding operator \(\tau\) \eqref{eq:tau}. These functions provide
two different ways of applying the threshold, ``uniform'' and
``dependent'' (\emph{i.e.}, the same threshold for each coefficient vs a
threshold normalized by the variance of each coefficient). The second
approach generally provides better results (especially when the weights
have been calculated via the frame). A comparative example of these two
approaches is given below (with \(\beta=2\) James-Stein attenuation
threshold).

\begin{Shaded}
\begin{Highlighting}[]
\NormalTok{diagWWt }\OtherTok{\textless{}{-}} \FunctionTok{colSums}\NormalTok{(}\FunctionTok{t}\NormalTok{(tf)}\SpecialCharTok{\^{}}\DecValTok{2}\NormalTok{)}
\NormalTok{thresh }\OtherTok{\textless{}{-}} \FunctionTok{sort}\NormalTok{(}\FunctionTok{abs}\NormalTok{(wcn))}
\NormalTok{opt\_thresh\_d }\OtherTok{\textless{}{-}} \FunctionTok{SURE\_MSEthresh}\NormalTok{(wcn, }
\NormalTok{                           wcf, }
\NormalTok{                           thresh, }
\NormalTok{                           diagWWt, }
                           \AttributeTok{beta=}\DecValTok{2}\NormalTok{, }
\NormalTok{                           sigma, }
                           \ConstantTok{NA}\NormalTok{,}
                           \AttributeTok{policy =} \StringTok{"dependent"}\NormalTok{,}
                           \AttributeTok{keepwc =} \ConstantTok{TRUE}\NormalTok{)}

\NormalTok{opt\_thresh\_u }\OtherTok{\textless{}{-}} \FunctionTok{SURE\_MSEthresh}\NormalTok{(wcn, }
\NormalTok{                           wcf, }
\NormalTok{                           thresh, }
\NormalTok{                           diagWWt, }
                           \AttributeTok{beta=}\DecValTok{2}\NormalTok{, }
\NormalTok{                           sigma, }
                           \ConstantTok{NA}\NormalTok{,}
                           \AttributeTok{policy =} \StringTok{"uniform"}\NormalTok{,}
                           \AttributeTok{keepwc =} \ConstantTok{TRUE}\NormalTok{)}
\end{Highlighting}
\end{Shaded}

We can plot MSE risks and their SUREs estimates as a function of the
threshold parameter (assuming that \(\sigma\) is known).

\begin{Shaded}
\begin{Highlighting}[]
\FunctionTok{plot}\NormalTok{(thresh, opt\_thresh\_u}\SpecialCharTok{$}\NormalTok{res}\SpecialCharTok{$}\NormalTok{MSE,}
     \AttributeTok{type=}\StringTok{"l"}\NormalTok{, }\AttributeTok{xlab =} \StringTok{"t"}\NormalTok{, }\AttributeTok{ylab =} \StringTok{"risk"}\NormalTok{, }\AttributeTok{log=}\StringTok{"x"}\NormalTok{)}
\FunctionTok{lines}\NormalTok{(thresh, opt\_thresh\_u}\SpecialCharTok{$}\NormalTok{res}\SpecialCharTok{$}\NormalTok{SURE}\SpecialCharTok{{-}}\NormalTok{n}\SpecialCharTok{*}\NormalTok{sigma}\SpecialCharTok{\^{}}\DecValTok{2}\NormalTok{, }\AttributeTok{col=}\StringTok{"red"}\NormalTok{)}
\FunctionTok{lines}\NormalTok{(thresh, opt\_thresh\_d}\SpecialCharTok{$}\NormalTok{res}\SpecialCharTok{$}\NormalTok{MSE, }\AttributeTok{lty=}\DecValTok{2}\NormalTok{)}
\FunctionTok{lines}\NormalTok{(thresh, opt\_thresh\_d}\SpecialCharTok{$}\NormalTok{res}\SpecialCharTok{$}\NormalTok{SURE}\SpecialCharTok{{-}}\NormalTok{n}\SpecialCharTok{*}\NormalTok{sigma}\SpecialCharTok{\^{}}\DecValTok{2}\NormalTok{, }\AttributeTok{col=}\StringTok{"red"}\NormalTok{, }\AttributeTok{lty=}\DecValTok{2}\NormalTok{)}
\FunctionTok{legend}\NormalTok{(}\StringTok{"topleft"}\NormalTok{, }\AttributeTok{legend=}\FunctionTok{c}\NormalTok{(}\StringTok{"MSE\_u"}\NormalTok{, }\StringTok{"SURE\_u"}\NormalTok{,}
                           \StringTok{"MSE\_d"}\NormalTok{, }\StringTok{"SURE\_d"}\NormalTok{),}
       \AttributeTok{col=}\FunctionTok{rep}\NormalTok{(}\FunctionTok{c}\NormalTok{(}\StringTok{"black"}\NormalTok{, }\StringTok{"red"}\NormalTok{), }\DecValTok{2}\NormalTok{), }
       \AttributeTok{lty=}\FunctionTok{c}\NormalTok{(}\DecValTok{1}\NormalTok{,}\DecValTok{1}\NormalTok{,}\DecValTok{2}\NormalTok{,}\DecValTok{2}\NormalTok{), }\AttributeTok{cex =} \DecValTok{1}\NormalTok{)}
\end{Highlighting}
\end{Shaded}

\begin{figure}

{\centering \includegraphics{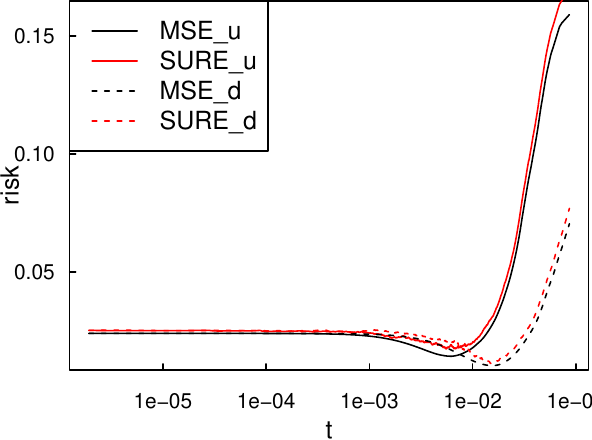} 

}

\caption{MSE risk and its SURE estimates as a function of the threshold parameter.}\label{fig:unnamed-chunk-29}
\end{figure}

Finally, the synthesis allows us to determine the resulting estimators
of \(f\), \emph{i.e.}, the ones that minimize the unknown MSE risks and
the ones that minimizes the SUREs.

\begin{Shaded}
\begin{Highlighting}[]
\NormalTok{wc\_oracle\_u }\OtherTok{\textless{}{-}}\NormalTok{ opt\_thresh\_u}\SpecialCharTok{$}\NormalTok{wc[, opt\_thresh\_u}\SpecialCharTok{$}\NormalTok{min[}\StringTok{"xminMSE"}\NormalTok{]]}
\NormalTok{wc\_oracle\_d }\OtherTok{\textless{}{-}}\NormalTok{ opt\_thresh\_d}\SpecialCharTok{$}\NormalTok{wc[, opt\_thresh\_d}\SpecialCharTok{$}\NormalTok{min[}\StringTok{"xminMSE"}\NormalTok{]]}
\NormalTok{wc\_SURE\_u }\OtherTok{\textless{}{-}}\NormalTok{ opt\_thresh\_u}\SpecialCharTok{$}\NormalTok{wc[, opt\_thresh\_u}\SpecialCharTok{$}\NormalTok{min[}\StringTok{"xminSURE"}\NormalTok{]]}
\NormalTok{wc\_SURE\_d }\OtherTok{\textless{}{-}}\NormalTok{ opt\_thresh\_d}\SpecialCharTok{$}\NormalTok{wc[, opt\_thresh\_d}\SpecialCharTok{$}\NormalTok{min[}\StringTok{"xminSURE"}\NormalTok{]]}

\NormalTok{hatf\_oracle\_u }\OtherTok{\textless{}{-}} \FunctionTok{synthesis}\NormalTok{(wc\_oracle\_u, tf)}
\NormalTok{hatf\_oracle\_d }\OtherTok{\textless{}{-}} \FunctionTok{synthesis}\NormalTok{(wc\_oracle\_d, tf)}
\NormalTok{hatf\_SURE\_u  }\OtherTok{\textless{}{-}} \FunctionTok{synthesis}\NormalTok{(wc\_SURE\_u, tf)}
\NormalTok{hatf\_SURE\_d  }\OtherTok{\textless{}{-}} \FunctionTok{synthesis}\NormalTok{(wc\_SURE\_d, tf)}

\NormalTok{res }\OtherTok{\textless{}{-}} \FunctionTok{data.frame}\NormalTok{(}\StringTok{"Input\_SNR"}\OtherTok{=}\FunctionTok{round}\NormalTok{(}\FunctionTok{SNR}\NormalTok{(f,tilde\_f),}\DecValTok{2}\NormalTok{),}
                  \StringTok{"MSE\_u"}\OtherTok{=}\FunctionTok{round}\NormalTok{(}\FunctionTok{SNR}\NormalTok{(f,hatf\_oracle\_u),}\DecValTok{2}\NormalTok{),}
                  \StringTok{"SURE\_u"}\OtherTok{=}\FunctionTok{round}\NormalTok{(}\FunctionTok{SNR}\NormalTok{(f,hatf\_SURE\_u),}\DecValTok{2}\NormalTok{),}
                  \StringTok{"MSE\_d"}\OtherTok{=}\FunctionTok{round}\NormalTok{(}\FunctionTok{SNR}\NormalTok{(f,hatf\_oracle\_d),}\DecValTok{2}\NormalTok{),}
                  \StringTok{"SURE\_d"}\OtherTok{=}\FunctionTok{round}\NormalTok{(}\FunctionTok{SNR}\NormalTok{(f,hatf\_SURE\_d),}\DecValTok{2}\NormalTok{))}
\end{Highlighting}
\end{Shaded}

\begin{table}[H]

\caption{\label{tab:risk}Comparison of SNR performance between  uniform and dependent policies.}
\centering
\begin{tabular}[t]{rrrrr}
\toprule
Input\_SNR & MSE\_u & SURE\_u & MSE\_d & SURE\_d\\
\midrule
\cellcolor{gray!6}{8.24} & \cellcolor{gray!6}{12.55} & \cellcolor{gray!6}{12.15} & \cellcolor{gray!6}{14.38} & \cellcolor{gray!6}{14.38}\\
\bottomrule
\end{tabular}
\end{table}

It can be seen from Table \ref{tab:risk} that in both cases, SURE
provides a good estimator of the MSE and therefore the resulting
estimators have performances close (in terms of SNR) to those obtained
by minimizing the unknown risk.

Equivalently, estimators can be obtained by the inverse of the SGWT
given by the function \texttt{inverse\_sgwt}. For exemple:

\begin{Shaded}
\begin{Highlighting}[]
\NormalTok{hatf\_oracle\_u }\OtherTok{\textless{}{-}} \FunctionTok{inverse\_sgwt}\NormalTok{(wc\_oracle\_u,}
\NormalTok{                              evalues, evectors, b)}
\end{Highlighting}
\end{Shaded}

Or if the coefficients have not been stored for each threshold value
(\emph{i.e.}, with the argument ``keepwc=FALSE'' when calling
\texttt{SUREthresh}) using the thresholding function
\texttt{beta\_thresh}, \emph{e.g.},

\begin{Shaded}
\begin{Highlighting}[]
\NormalTok{wc\_oracle\_u }\OtherTok{\textless{}{-}} \FunctionTok{betathresh}\NormalTok{(wcn, }
\NormalTok{                          thresh[opt\_thresh\_u}\SpecialCharTok{$}\NormalTok{min[[}\DecValTok{1}\NormalTok{]]], }\DecValTok{2}\NormalTok{)}
\end{Highlighting}
\end{Shaded}

Notably, SURE can also be applied in a level-dependent manner using
\texttt{SUREthresh} at each scale (the output of \texttt{SUREthresh} can
be retrieve with the argument ``keepSURE = TRUE'' at the function call).

\begin{Shaded}
\begin{Highlighting}[]
\NormalTok{J }\OtherTok{\textless{}{-}} \FunctionTok{floor}\NormalTok{(}\FunctionTok{log}\NormalTok{(lmax)}\SpecialCharTok{/}\FunctionTok{log}\NormalTok{(b)) }\SpecialCharTok{+} \DecValTok{2}
\NormalTok{LD\_opt\_thresh\_u }\OtherTok{\textless{}{-}} \FunctionTok{LD\_SUREthresh}\NormalTok{(}\AttributeTok{J=}\NormalTok{J, }
                             \AttributeTok{wcn=}\NormalTok{wcn, }
                             \AttributeTok{diagWWt=}\NormalTok{diagWWt, }
                             \AttributeTok{beta=}\DecValTok{2}\NormalTok{, }
                             \AttributeTok{sigma=}\NormalTok{sigma,}
                             \AttributeTok{hatsigma=}\ConstantTok{NA}\NormalTok{,}
                             \AttributeTok{policy =} \StringTok{"uniform"}\NormalTok{,}
                             \AttributeTok{keepSURE =} \ConstantTok{FALSE}\NormalTok{)}
\NormalTok{hatf\_LD\_SURE\_u }\OtherTok{\textless{}{-}} \FunctionTok{synthesis}\NormalTok{(LD\_opt\_thresh\_u}\SpecialCharTok{$}\NormalTok{wcLDSURE, tf)}
\FunctionTok{print}\NormalTok{(}\FunctionTok{paste0}\NormalTok{(}\StringTok{"LD\_SURE\_u = "}\NormalTok{,}\FunctionTok{round}\NormalTok{(}\FunctionTok{SNR}\NormalTok{(f,hatf\_LD\_SURE\_u),}\DecValTok{2}\NormalTok{),}\StringTok{"dB"}\NormalTok{))}
\CommentTok{\#\textgreater{} [1] "LD\_SURE\_u = 13.1dB"}
\end{Highlighting}
\end{Shaded}

Even though the SURE no longer depends on the original signal, it does
depend on \(\sigma^2\), two naive (biased) estimators are obtained via
\texttt{GVN} or \texttt{HPVN} functions (see \citet{de2019data} for more
details). Another possible improvement would be to use a scale-dependent
variance estimator (especially in the case of ``policy
=''dependent''\,``).

Furthermore, the major limitations are the need to diagonalize the
graph's Laplacian, and the calculation of the weights involved in the
SURE (which requires an explicit calculation of the frame). To address
the first limitation, several strategies have been proposed in the
literature, notably via approximation by Chebyshev polynomials (see
\citet{hammond2011wavelets} or \citet{shuman18}). Combined with these
approximations, a Monte Carlo method to estimate the SURE weights has
been proposed in \citet{chedemail22}, extending the applicability of
SURE to large graphs.

\bibliography{references}

\end{document}